\documentclass[8.4pt,twocolumn]{article}
\usepackage{graphicx}
\usepackage{amssymb}
\usepackage{enumerate}
\usepackage{layout}

\newcommand{\be}{\begin{equation}}
\newcommand{\ee}{\end{equation}}
\newcommand{\n}{\noindent}
\newcommand{\dd}{\,{\rm d}}

\begin{document}

\title{\large \bf Cooperative Gating and Spatial Organization of Membrane Proteins through Elastic Interactions}   
\author{
\small Tristan Ursell$^{1}$, Kerwyn Huang$^{2}$, Eric Peterson$^{3}$ and Rob Phillips$^{1,4}$\footnote{corresponding author:  phillips@pboc.caltech.edu}  \\ \small Departments of $^{1}$Applied Physics and $^{3}$Physics, California Institute of Technology, Pasadena, CA 91125, USA\\ \small $^{4}$Kavli Nanoscience Institute, Pasadena, CA 91125, USA\\\small $^{2}$Department of Molecular Biology, Princeton University, Princeton, NJ 08544, USA\\
\small Submitted to {\it PLoS Computational Biology}\\
}
\date{\small\today} 
\maketitle
 
\n
{\footnotesize Biological membranes are elastic media in which the presence of a transmembrane protein leads to local bilayer deformation.  The energetics of deformation allow two membrane proteins in close proximity to influence each other's equilibrium conformation via their local deformations, and spatially organize the proteins based on their geometry.  We use the mechanosensitive channel of large conductance (MscL) as a case study to examine the implications of bilayer-mediated elastic interactions on protein conformational statistics and clustering.  The deformations around MscL cost energy on the order of $10\,k_BT$ and extend $\sim3\,\mbox{nm}$ from the protein edge, as such elastic forces induce cooperative gating and we propose experiments to measure these effects.  Additionally, since elastic interactions are coupled to protein conformation, we find that conformational changes can severely alter the average separation between two proteins.  This has important implications for how conformational changes organize membrane proteins into functional groups within membranes. 
}\\

\n
{\bf\large Introduction}

Biological membranes are active participants in the function and spatial organization of membrane proteins \cite{Mouritsen1993,Lee2003,Jensen2004}.  At the simplest level, the membrane positions proteins into a two-dimensional space, where they are often laterally organized into groups. These groups can serve specific purposes on the cell surface and within organelles, such as sensing, adhesion and transport  \cite{Bray1998,Sourjik2004,Gibbs2004,Vale2005,Freund2005,Engelman2005}.  Electrostatic and van der Waal's forces help drive lateral organization \cite{Goulian1993}, however there is an additional class of purely bilayer-mediated elastic forces that can facilitate the formation of complexes of membrane proteins.  

Conformational changes of membrane proteins result from a wide range of environmental factors including temperature, pH, ligand and small molecule binding, membrane voltage, and membrane tension.  Likewise, conformational state is often tightly coupled with function ({\it e.g.} for ion channels) \cite{SukharevAnnRev1997,Clapham2001,Barry2005}. In this work, we demonstrate how elastic interactions can communicate information about protein conformation from one neighboring protein to another coupling their conformational state.  Additionally, we find that these interactions lead to spatial organization within the bilayer that is strongly dependent on protein conformation.

We suggest that elastic forces play a role in the function and spatial organization of many membrane proteins across many cell types, given the generically high areal density of membrane proteins \cite{Engelman2004} and the strength of these interactions.  We use the mechanosensitive channel of large conductance (MscL) from {\it E. coli} as the model protein for this study. MscL is a transmembrane homo-pentamer found in the plasma membrane of {\it E. coli} (and many other bacteria) serving as an emergency relief valve under hypo-osmotic shock \cite{SukharevAnnRev1997,Rees1998,Pivetti2003}.  As membrane tension increases, this non-selective ion channel changes conformation from a closed state to an open state, releasing water and osmolytes \cite{SukharevJGP1999,SukharevNature2001}.  Though several substates have been identified in this gating transition, the relatively short dwell-times in these substates as compared to the fully open or fully closed states, allows us to approximate the protein as a simple two-state system \cite{SukharevJGP1999,SukharevBJ2004}.  Crystal and  electron-paramagnetic-resonance structures suggest the bilayer-spanning region is nearly cylindrical in both the open and closed conformations \cite{Rees1998,PerozoJGP2001,PerozoNature2002}, making MscL particularly amenable to mechanical modeling.  Electrophysiology of reconstituted channels allows measurement of the state of one or more of these proteins with excellent temporal and number resolution.  Therefore, theoretical predictions for how elastic interactions change the gating behavior of a MscL protein can be readily tested using electrophysiology and other experimental techniques.

\begin{figure}
\begin{center}
\includegraphics[width=3.4in]{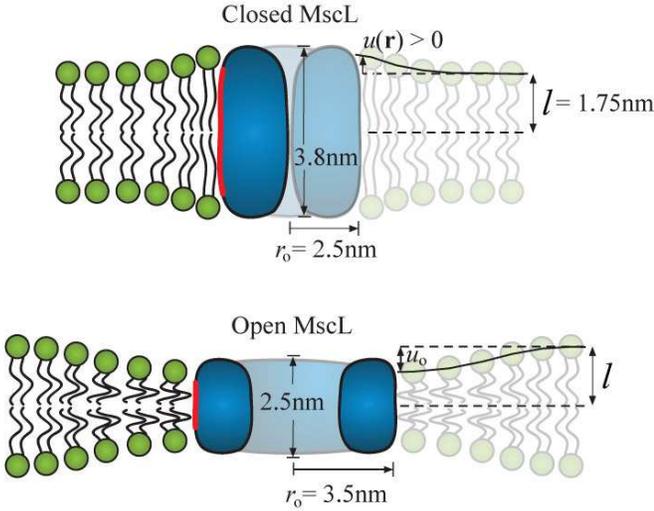}
\caption{ \small Schematic of Bilayer Deformations due to MscL.
Mismatch between the hydrophobic regions of the lipid bilayer and an integral membrane protein gives rise to bending and compression deformations in each leaflet of the bilayer.  The largest deformations occur at the protein-lipid interface, and over the scale of a few nanometers the bilayer returns to its unperturbed state.  MscL is shown schematically at zero tension in its closed and open states with relevant dimensions.  The red region of the protein indicates the hydrophobic zone.  The hydrophobic mismatch at the protein-lipid interface is denoted by $u_o$. The deformation profile, denoted by $u(\bf{r})$, is measured with reference to the unperturbed leaflet thickness ($l$) from the protein center at ${\bf r}=0$. }
\label{Fig1}
\end{center}
\end{figure}

Following earlier work, we use continuum mechanics to break down the deformation caused by a cylindrical transmembrane protein into a term penalizing changes in bilayer thickness and a term penalizing bending of a bilayer leaflet \cite{Huang1986,Dan1993,Dan1996,AndersenBJ1998,Wiggins2005}, and we introduce a third term that preserves bilayer volume under deformation \cite{Sachs2004}.  Due to its structural symmetry, MscL can be characterized by its radius and bilayer-spanning thickness in its two conformations ({\it i.e.} open and closed), neglecting any specific molecular detail (see Figure \ref{Fig1}).  As these geometric parameters change with conformation, the bilayer-mediated interaction between two channels is altered.  Using the interaction potentials in each combination of conformations, we explore how both the single-channel and interacting energetics affect the spatial and conformational behavior of two channels.

In the first section we cover the physical principles behind bilayer deformation due to the presence of membrane proteins.  In the second section we explore the differences in gating behavior of two MscL proteins when held at a fixed separation. In the third section we explore the conformational and spatial behavior of diffusing MscL proteins as a function of areal density. Finally, in the fourth section we discuss the relevance of these forces as compared to other classes of bilayer-mediated forces and support our hypotheses with results from previous experiments.\\

\n
{\bf\large Results}\\

\n
{\large Elastic Deformation Induced by Membrane Proteins}

The bilayer is composed of discrete lipid molecules whose lateral diffusion ($D\sim10\,\mu\mbox{m}^2/s$) \cite{Kahya2003} is faster than the diffusion of transmembrane proteins ($D\sim0.1-1\mu\mbox{m}^2/s$) \cite{Doeven2005,Gambin2006,Guigas2006}.  In the time it takes a transmembrane protein to diffuse one lipid diameter, many lipids will have exchanged places near the protein to average out the discreteness of the lipid molecules.  Additionally, the transition time for protein conformational change ($\sim5\,\mu s$) \cite{Lester2004} is slow compared to lipid diffusion.  Hence, we argue the bilayer can be approximated as a continuous material in equilibrium with well-defined elastic properties \cite{HuangBJ1999}.  Further, we choose to formulate our analysis in the language of continuum mechanics, rather than  lateral pressure profiles \cite{Cantor1999}. In particular, each leaflet of the bilayer resists changes in the angle between adjacent lipid molecules, leading to bending stiffness of the bilayer \cite{Huang1986,Helfrich1973}.  Likewise, the bilayer has a preferred spacing of the lipid molecules in-plane and will resist any changes in this spacing due to external tension \cite{Evans2000}.  Finally, experiments suggest that the volume per lipid is conserved \cite{Tosh1986,Seemann2003} such that changes in bilayer thickness are accompanied by changes in lipid spacing \cite{Lee2003,HuangBJ1999}.

Transmembrane proteins can compress and bend a bilayer leaflet via at least two mechanisms.  The protein can force the bilayer to adopt a new thickness, matching the hydrophobic region of the protein to the hydrophobic core of the bilayer.  Additionally, a non-cylindrical protein can induce a slope in the leaflet at the protein-lipid interface \cite{AndersenBJ1998,Dan1998}.

For transmembrane proteins like MscL, that can be approximated as cylindrical, symmetry dictates that the deformation energy of the bilayer is twice the deformation energy of one leaflet. Presuming the protein does not deform the bilayer too severely, we can write the bending and compression (thickness change) energies in a form analogous to Hooke's law, and account for external tension with a term analogous to $PV$ work.  We denote the deformation of the leaflet by the function $u({\bf r})$, which measures the deviation of the lipid head-group from its unperturbed height as a function of the position $\bf{r}$ (see Figure \ref{Fig1}).  In all the calculations that follow the physical parameters chosen are representative of a typical phosphatidylcholine (PC) lipid bilayer and the number of lipids in this model bilayer is fixed. The energy penalizing compression of the bilayer is 
\be{
G_{\mbox{\tiny comp}}=\frac{K_A}{2}\int\left(\frac{u({\bf r})}{l}\right)^2\dd^2{\bf r},
}\ee

\n
where $K_A$ is the bilayer area stretch modulus ($\sim58\,k_BT/\mbox{nm}^2$, $k_BT$ is the thermal energy unit) and $l$ is the unperturbed leaflet thickness ($\sim1.75\,\mbox{nm}$) \cite{Evans2000}.  The bending energy of a leaflet is
\be{
G_{\mbox{\tiny bend}}=\frac{\kappa_b}{4}\int(\nabla^2 u({\bf r})-c_o)^2\dd^2{\bf r},
}\ee

\n
where $\kappa_b$ ($\sim14\,k_BT$) is the bilayer bending modulus, $c_o$ is the spontaneous curvature of the leaflet \cite{HuangBJ1999,Evans2000,Helfrich1995}, and $\nabla^2=\partial^2/\partial x^2+\partial^2/\partial y^2$ is the Laplacian operator.  

Coupling external tension to bilayer deformations is more subtle than the previous two energetic contributions.  We note that the bilayer is roughly forty times more resistant to volume change than area change \cite{Tosh1986,Seemann2003}, hence if a transmembrane protein locally thins the bilayer, lipids will expand in the area near the protein to conserve volume.  Likewise, if the protein locally thickens the bilayer,  lipids near the protein will condense (see Figure \ref{Fig1}).  Therefore, the area change near the protein is proportional to the compression $u(\bf{r})$, and the work done on the bilayer is the integrated area change multiplied by tension
\be{
G_{\mbox{\tiny ten}}=\tau\int\frac{u({\bf r})}{l}\dd^2{\bf r},
}\ee

\n
where $\tau$ is the externally applied bilayer tension.  Slightly below bilayer rupture, and near the expected regime of MscL gating, $\tau\simeq 2.6\,k_BT/\mbox{nm}^2$ \cite{SukharevJGP1999,Evans2000}.  In total, the bilayer deformation energy is
\be{
G=\frac{1}{2}\int\left(K_A\left(\frac{u}{l}+\frac{\tau}{K_A}\right)^2+\kappa_b\left(\nabla^2u-c_o\right)^2\right)\dd^2{\bf r},
}\ee

\n
where we have made use of the constant bilayer area to elucidate the interplay between tension and compression\footnote{Specifically, we added a constant proportional to membrane area and $\tau^2$, which is identically zero when calculating differences in free energy.}. 

To obtain the length and energy scales of these deformations, we non-dimensionalize the bilayer deformation energy, $G$.  We scale both the position ${\bf r}$ and displacement $u({\bf r})$ by $\lambda=(\kappa_bl^2/K_A)^{1/4}\simeq1\,\mbox{nm}$, the natural length scale of deformation, to give the new variables $\mbox{\boldmath$\rho$}$ and $\eta(\mbox{\boldmath$\rho$})$ respectively, where $\mbox{\boldmath$\rho$}={\bf r}/\lambda$ and $\eta(\mbox{\boldmath$\rho$})=u({\bf r})/\lambda$.  Then $G$ can be written as
\be{
G=\frac{\kappa_b}{2}\int\left(\left(\eta+\chi\right)^2+\left(\nabla^2\eta-\nu_o\right)^2\right)\dd^2\mbox{\boldmath$\rho$},
\label{eqG}
}\ee

\n
where $\nu_o=\lambda c_o$ is the dimensionless spontaneous curvature and $\chi=\tau l/K_A\lambda$ is the dimensionless tension, which is $\simeq0.09$ in the regime of MscL gating .  The energy scale is set by the bending modulus, $\kappa_b$.

Using the standard Euler-Lagrange equation from the calculus of variations \cite{Arfken}, the functional for the deformation energy can be translated into the partial differential equation \footnote{With no loss of generality, the equation that governs deformation can also be written in the parameter free form $\nabla^4\eta+\eta=0$, with all parametric sensitivity absorbed into the boundary conditions.}
\be{
\nabla^4\eta+\eta+\chi=0.
\label{pdeeq}
}\ee

\n
The deformation profile $u(\bf{r})$ that solves this partial differential equation depends on four boundary conditions.  In the far-field, we expect the bilayer to be flat and slightly thinner in accordance with the applied tension, {\it i.e.} $|\nabla u(\infty)|=0$ and $u(\infty)=-\tau l / K_A$, respectively. At the protein-lipid interface (${\bf r}=r_o$) the hydrophobic regions of the protein and the bilayer must be matched, {\it i.e.} $u(r_o)=u_o$ (see Figure \ref{Fig1}), where $u_o$ is one-half the mismatch between the hydrophobic region of the protein and the hydrophobic core of the bilayer.  Finally, the slope of the bilayer at the protein-lipid interface is set to zero ({\it i.e.} $|\nabla u(r_o)|=0$).  The motivation for this last boundary condition is subtle and will be examined in more detail in the Discussion.  

To understand how the deformation energy scales with hydrophobic mismatch ($u_o$), protein size ($r_o$), and tension ($\tau$), we solve Eq.~\ref{pdeeq} analytically  for a single cylindrical protein. The deformation energy is
\be{
G_{\mbox{\tiny single}}=\pi\kappa_b\int_{\rho_o}^{\infty}\left(\left(\eta+\chi\right)^2+\left(\nabla^2\eta-\nu_o\right)^2\right)\rho\dd\rho,
}\ee

\n
where $\rho_o=r_o/\lambda$ is the dimensionless radius of the protein.  The bilayer deformation around a single protein is a linear combination of zeroth order modified Bessel functions of the second kind ($K_0$) \cite{AndersenBJ1998,Wiggins2005}.  For proteins such as MscL with a radius larger than $\lambda$ ({\it i.e.} $1\,\mbox{nm}$) the deformation energy is well-approximated by
\be{
G_{\mbox{\tiny single}}=\pi\kappa_b\left(\frac{u_o}{\lambda}+\frac{\tau}{K_A}\frac{l}{\lambda}\right)^2\left(1+\sqrt{2}\frac{r_o}{\lambda}\right).
\label{eqsingle}
}\ee

\n
The deformation energy scales linearly with protein radius and depends quadratically on the {\it combination} of hydrophobic mismatch ($u_o$) and tension ($\tau$).  This makes the overall deformation energy particularly sensitive to the hydrophobic mismatch, and hence leaflet thickness $l$.  The deformation energy is fairly insensitive to changes in $K_A$ ({\it i.e.} most terms in the energy are sublinear), and generally insensitive to changes in the bending modulus since $G\propto\kappa_b^{1/4}$.  

Using our standard elastic bilayer parameters and the dimensions of a MscL channel (see Figure \ref{Fig1}), the change in deformation energy between the closed and open states is $\Delta G_{\mbox{\tiny single}}\simeq50\,k_BT$.  The measured value for the free energy change of gating a MscL protein, including internal changes of the protein and deformation of surrounding lipids is $\simeq51\,k_BT$ \cite{SukharevBJ2004}.  This close correspondence does not indicate that bilayer deformation accounts for all of the free energy change of gating \cite{Yoshimura1999}, but does suggest that it is a major contributor.

The gating energy of two channels in close proximity is a complex function of their conformations and the distance between them.  As two proteins come within a few nanometers of each other ({\it i.e.} a few $\lambda$), the deformations which extend from their respective protein-lipid interfaces begin to overlap and interact.  The bilayer adopts a new shape ({\it i.e.} a new $u(\bf{r})$), distinct from the deformation around two independent proteins, and hence the total deformation energy changes as well.  This is the physical origin of the elastic interaction between two bilayer-deforming proteins \cite{Dan1993,Dan1996}.

\begin{figure}
\begin{center}
\includegraphics[width=3.5in]{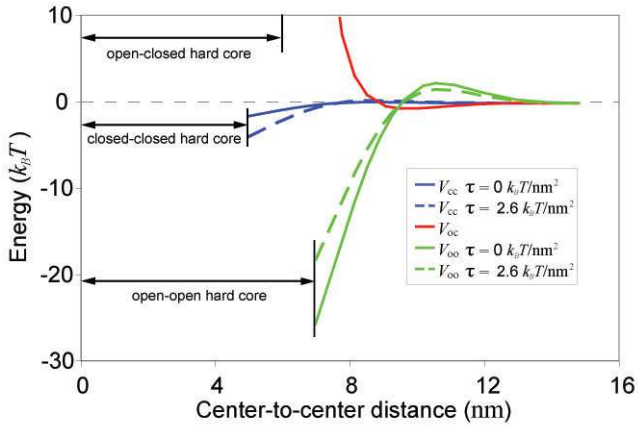}
\caption{ \small Elastic Potentials between Two MscL Proteins. To minimize deformation energy, two transmembrane proteins exert elastic forces on each other.  MscL has three distinct interaction potentials between its two distinct conformations.  External tension weakens the interaction between two open channels ($V_{\mbox{\tiny oo}}$) and strengthens the interaction between two closed channels ($V_{\mbox{\tiny cc}}$), but has almost no effect on the interaction between an open and closed channel ($V_{\mbox{\tiny oc}}$).  The open-open and closed-closed interactions are both more strongly attracting than the open-closed interaction, indicating that elastic potentials favor interactions between channels in the same state. The `hard core' distance is where the proteins' edges are in contact.}
\label{Fig2}
\end{center}
\end{figure}

Each protein imposes its own local boundary conditions on the bilayer, that vary with conformation, hence the deformation around a pair of proteins is a function of their individual conformation and the distance between them.  A MscL protein has two distinct conformations, hence there can be pairwise interactions between two closed channels, an open and a closed channel, or two open channels (see Figure \ref{Fig2}).  Tension also affects the deformations.  The hydrophobic mismatch can be either positive or negative ({\it i.e.} the protein can be thicker or thinner than the bilayer), thus tension will strengthen the interaction of proteins that are thicker than the bilayer ({\it e.g.} the closed-closed interaction of two MscL proteins) and weaken the interaction of proteins that are thinner than the bilayer ({\it e.g.} the open-open interaction). This effect is demonstrated in Figure \ref{Fig2}.  The interactions due to leaflet deformations have been explored before \cite{Dan1993,Dan1996}, but our model elucidates the role that these interactions can play in communicating conformational information between proteins.  Additionally, in our model, tension can play an important role in determining the overall deformation energy around a protein.  

In a one-dimensional model, the interaction potentials can be solved for analytically.  For two identical proteins in close proximity ({\it e.g.} closed-closed and open-open interactions), the approximate shape of the potential is linearly attractive $\kappa_b(u_o/\lambda)^2(d/2\lambda-\sqrt{2})$.  Between two dissimilar proteins in close proximity ({\it e.g.} open-closed interaction) the potential is approximately $\kappa_b\lambda u_o^2\pi^4/4d^3$, where in both cases $d$ is measured from the edges of the proteins.  This illustrates the general principle that two similar proteins attractively interact, while two dissimilar proteins tend to repel each other.  In a two-dimensional bilayer, however, the geometry of the two proteins makes it difficult to solve for the interaction analytically, thus numerical techniques were used (see Materials and Methods).  

This theoretical framework provides a strong foundation for understanding how protein geometries and lipid properties give rise to elastic interactions.  With this, we can investigate how elastic forces change the conformational statistics of a two-state protein population.\\

\n
{\large Gating Behavior of Two Interacting Channels}

To probe the range of separations over which elastic interactions affect the gating of two MscL proteins, we need to account for the non-interacting energetics of gating a single channel in addition to the interactions between two channels.  The non-interacting energy is the sum of three effects.  First, there is some energetic cost to deform the surrounding membrane, which we already calculated as $\Delta G_{\mbox{\tiny single}}$.  Second, there is some cost to change the protein's internal conformation, independent of the membrane.  Together, these first two effects are the gating energy $\Delta G_{\mbox{\tiny gate}}\simeq51\,k_BT$ for MscL \cite{SukharevBJ2004}.  Finally, there is an energetic mechanism that overcomes these costs and opens the channel as tension increases.  This mechanism is provided by the bilayer tension working in concert with the conformational area change of the protein ($\Delta A\simeq20\,\mbox{nm}^2$ for MscL \cite{SukharevBJ2004}).  Given the experimentally determined values for $\Delta G_{\mbox{\tiny gate}}$ and the area change during gating, the critical tension, defined by $\Delta G_{\mbox{\tiny gate}}=\tau_c\Delta A$, is $\tau_c=2.6\,k_BT/\mbox{nm}^2$.

In our thermodynamic treatment we need to keep track of the conformations of each protein in a population in a way that allows us to tabulate the non-interacting and interacting contributions to the free energy. To this end, we assign a state variable, $s_i$, to each channel indicating the conformational state of a protein, where $s_{i}=0$ indicates that the $i$th channel is closed and $s_{i}=1$ indicates that the $i$th channel is open.  The non-interacting energy for two channels is then
\be{
H_{\mbox{\tiny non}}(s_1,s_2;\tau)=\left(\Delta G_{gate}-\tau\Delta A\right)(s_1+s_2).
}\ee

\n
If both channels are closed ($s_1=s_2=0$) the free energy is defined to be zero.  If one channel is open and the other closed ($s_1=1,s_2=0$ or $s_1=0,s_2=1$) this counts as the cost to gate one channel working against the benefit at a particular tension to opening the channel.  Likewise, this counts twice if both channels are open ($s_1=s_2=1$). We will measure all energies that follow in units of $k_BT$ ($\simeq4.14\times10^{-21}J$).

As we alluded to earlier, the interacting component of the free energy between two channels is a function of their states ($s_1$ and $s_2$), their edge separation ($d$), and the tension.  Using a numerical relaxation technique to minimize the functional in Eq. \ref{eqG} (see Materials and Methods), we calculated the interaction potentials $H_{\mbox{\tiny int}}(s_1,s_2,d; \tau)$ for a range of tensions and separation distances (see Figure \ref{Fig2}).  The total energy, $H_{\mbox{\tiny non}}+H_{\mbox{\tiny int}}$, is used to derive the Boltzmann weight for the three possible configurations of the two-channel system,
\be{
z(s_1,s_2)=e^{-\left(H_{\mbox{\tiny non}}(s_1,s_2;\tau)+H_{\mbox{\tiny int}}(s_1,s_2,d;\tau)\right)}.
}\ee

\n
The probability that the system has two closed channels is
\be{
P_0=\frac{z(0,0)}{Z},
}\ee

\n
where the partition function $Z$ is the sum of the Boltzmann weights for all possible two-channel configurations,
\be{
Z=\sum_{s_1,s_2=0}^{1}z(s_1,s_2)=z(0,0)+2z(0,1)+z(1,1).
}\ee

\n
Likewise, the probabilities for the system to have exactly one or two open channels are
\be{
P_{1}=\frac{2z(0,1)}{Z}\,\,\,\,\,\,{\rm and}\,\,\,\,\,\,P_{2}=\frac{z(1,1)}{Z},
}\ee

\n
respectively.  Finally, the probability for any one channel in this two channel system to be open is
\be{
P_{\mbox{\tiny open}}(\tau,d)=\frac{z(0,1)+z(1,1)}{Z}.
}\ee

\begin{figure}
\begin{center}
\includegraphics[width=3.4in]{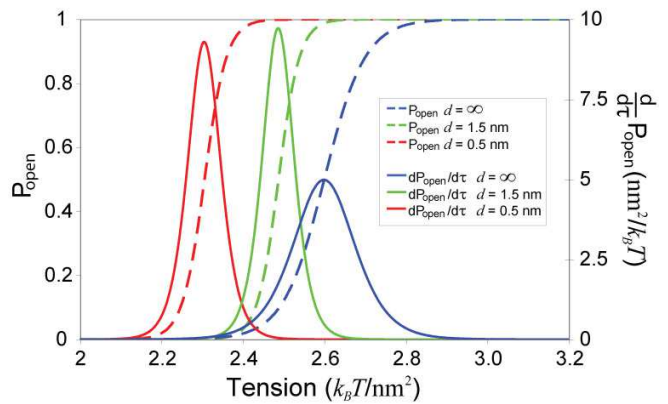}
\caption{ \small Conformational Statistics of Interacting MscL Proteins.  Interactions between neighboring channels lead to shifts in the probability that a channel will be in the open state (dashed lines).  The sensitivity and range of response to tension, ${\mbox{d}P_{\mbox{\tiny open}}/\mbox{d}\tau}$, are also affected by bilayer deformations (solid lines).  ${P_{\mbox{\tiny open}}}$ and ${\mbox{d}P_{\mbox{\tiny open}}/\mbox{d}\tau}$ are shown for separations of $0.5\,\mbox{nm}$ (red) and $1.5\,\mbox{nm}$ (green) with reference to non-interacting channels at $d=\infty$ (blue).  Interactions shift the critical gating tension for the closest separation by $\sim12\%$.  Additionally, the peak sensitivity is increased by $\sim90\%$ from $\sim5\mbox{nm}^2/k_BT$  to $\sim9.5\mbox{nm}^2/k_BT$, indicating a Hill coefficient of $\sim2$.}
\label{Fig3}
\end{center}
\end{figure}

\n
If the distance between two channels is much greater than $\lambda$, they will behave independently.  As the channels get closer ($d\lesssim5\lambda$) they begin to interact and their conformational statistics are altered.  $P_{\mbox{\tiny open}}$ as a function of tension for certain fixed separations is shown in Figure \ref{Fig3}.  The open-open interaction is the most energetically favorable for most separations, hence the transition to the open state generally shifts to lower tensions as the distance between the two proteins is decreased.  

Interactions also affect channel `sensitivity', defined as the derivative of $P_{\mbox{\tiny open}}$ with respect to tension, which quantifies how responsive the channel is to changes in tension.  The full-width at half maximum of this peaked function is a measure of the range of tension over which the channel has an appreciable response.  The area under the sensitivity curve is equal to 1, hence increases in sensitivity are always accompanied by decreases in range of response, as demonstrated by the effects of the beneficial open-open interaction on channel statistics (see Figure \ref{Fig3}).

In summary, we find that elastic interactions between two proteins have significant effects when the protein edges are closer than $\sim5\,\mbox{nm}$.  At these separations the elastic interactions alter the critical gating tension and change the tension sensitivity of the channel (see Figure \ref{Fig3}).  The critical gating tension and sensitivity are the key properties which define the transition to the open state, and are analogs to the properties which define the transition of {\it any} two-state ion channel.  Hence we have shown that elastic interactions can affect channel function at a fundamental level.\\

\n
{\large Interactions between Diffusing Proteins}

With an understanding of how two proteins will interact at a fixed distance, we now study the conformational statistics of two freely-diffusing MscL proteins allowed to interact via their elastic potentials. In biological membranes, transmembrane proteins that are not rigidly attached to any cytoskeletal elements are often free to diffuse throughout the membrane and interact with various lipid species as well as other membrane proteins.  On average, the biological areal density of such proteins is high enough ($\sim100-1000\,\mbox{nm}^2$/protein \cite{Engelman2004}) that elastic interactions should alter the conformational statistics and average protein separations.

We expect that if two MscL proteins are diffusing and interacting, the open probability will be a function of their areal density as well as the tension.  It then follows that for a given areal density, elastic interactions will couple conformational changes to the average separation between the proteins. To calculate the open probability of two diffusing MscL proteins, the Boltzmann weight for these proteins to be in the conformations $s_1$ and $s_2$ must be summed at every possible position, giving
\be{
\left<z(s_1,s_2)\right>=e^{-H_{\mbox{\tiny non}}}\int\int e^{-H_{\mbox{\tiny int}}}\dd^2{\bf r}_1\dd^2{\bf r}_2,
\label{eq17}
}\ee

\n
where $\left<\ldots\right>$ indicates a sum over all positions.  The distance between the proteins is measured center-to-center as $|{\bf r}_1-{\bf r}_2|$ and only the absolute distance between the two proteins determines their interaction, hence we can rewrite the integrand as a function of $r=|{\bf r}_1-{\bf r}_2|$.  We then change the form of the integrand to
\be{
e^{-H_{\mbox{\tiny int}}(s_1,s_2,r;\tau)}=1+f_{12}(r),
\label{eq18}
}\ee

\n
which allows us to separate the interacting effects from the non-interacting effects (the function $f_{12}$ is often called the Mayer-$f$ function).  Thus, the position-averaged Boltzmann weights are
\be{
\left<z(s_1,s_2)\right>=e^{-H_{\mbox{\tiny non}}}\left(1+\frac{2\pi}{A}\int_0^{\infty}f_{12}(r)r\dd r\right),
\label{eqV}
}\ee

\n
where $A$ is the total area occupied by the two proteins.  Following our previous calculations, the probability that any one channel is open in this two-channel system is
\be{
P_{\mbox{\tiny open}}(\tau,\alpha)=\frac{\left<z(0,1)\right>+\left<z(1,1)\right>}{\left<Z\right>},
}\ee

\n
where $\alpha$ is the area per protein ({\it i.e.} $\alpha=A/2$) and $\left<Z\right>=\sum_{s_1,s_2}\left<z(s_1,s_2)\right>$.  

\begin{figure}
\begin{center}
\includegraphics[width=3.3in]{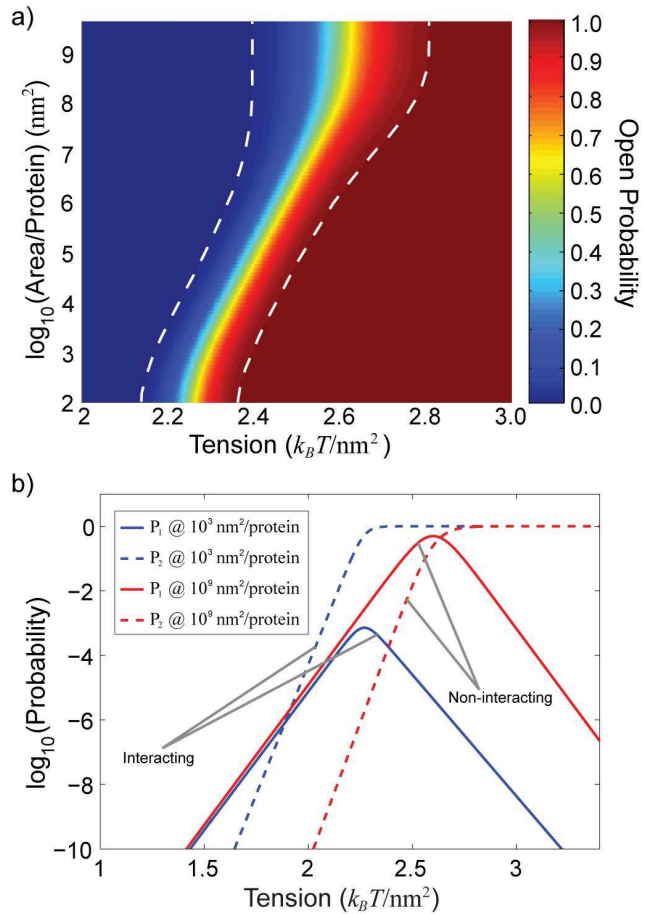}
\caption{ \small Elastic Interactions Lower Open Probability Transition and Couple Conformation Changes.
Two MscL proteins in a square box of area $A$ diffuse and interact via their elastic potentials.  a) At high area per protein, the response to tension is the same as an independent channel.  As the area per protein decreases, the more beneficial open-open interaction (see Figure \ref{Fig2}) shifts the open probability to lower tensions and decreases the range of response (dashed lines) while increasing the peak sensitivity, indicating that areal density can alter functional characteristics of a transmembrane protein. b) The probability for exactly one channel to be open ($P_1$ - solid lines) is shown at a high (red) and low (blue) area per protein.  For tensions past the critical tension, interacting channels are $\sim1000$ times less likely to gate individually.  The probability for both channels to be open simultaneously ($P_2$ - dashed lines) is shown for high (red) and low (blue) area per protein.  The tension at which two simultaneously open channels are favored is significantly lower for interacting channels.  Together these facts signify a tight coupling of the conformational changes for two interacting channels. }
\label{Fig4}
\end{center}
\end{figure}

In Figure \ref{Fig4}a, we plot $P_{\mbox{\tiny open}}(\tau,\alpha)$ over a wide range of areal density, from the area of $\sim100$ lipids up to areas on the whole-cell scale.  The more beneficial open-open interaction tends to shift the transition to the open state to lower tensions, with the most pronounced effect when the two proteins are most tightly confined.  For the estimated biological membrane protein density of $\sim10^2-10^3\,\mbox{nm}^2$/protein \cite{Engelman2004}, the gating tension is decreased by $\sim13\%$, the sensitivity is increased by $\sim85\%$ and the range of response is decreased by $\sim55\%$.  For the {\it in vivo} expression of MscL of $\sim10^5-10^6\,\mbox{nm}^2$/protein \cite{Stokes2003} the gating tension is reduced by $\sim7\%$, the sensitivity is increased by $\sim70\%$ and the range of response is decreased by $\sim40\%$.  These changes in gating behavior are accessible to electrophysiological experiments where MscL proteins can be reconstituted at a known areal density ($\sim10^5-10^7\,\mbox{nm}^2/$protein), and the open probability can be measured as a function of tension.  

In addition to lowering the critical tension and augmenting channel sensitivity, the conformational states of channels are tightly coupled by their interaction.  The probability that exactly one channel is open ($P_1$) decreases dramatically as areal density increases.  For tensions above the critical tension, interacting channels ($\sim10^3\,\mbox{nm}^2/$protein) are nearly three orders of magnitude less likely to gate as single channels than their non-interacting counterparts  ($\sim10^9\,\mbox{nm}^2/$protein), as shown in Figure \ref{Fig4}b.  Additionally, the tension at which it is more likely to have {\it both} channels open, rather than a single channel, is significantly lower for interacting channels, signaling that gating is a tightly coupled process.  In addition to altering the open probability of two channels, the favorable open-open interaction provides an energetic barrier to leaving the open-open state.  Based on a simple Arrhenius argument, the average open lifetime of two channels that are both open and interacting will be orders of magnitude longer than two open but noninteracting channels.

Having shown conformational coupling over a range of areal densities, it is reasonable to expect that elastic interactions will affect the separation between the two proteins. We ask, how do interactions affect the average separation between proteins?  How often will we find the two proteins separated by a distance small enough that we can consider them `dimerized'?

From Eqs.~\ref{eq17} and \ref{eq18} it follows that the Boltzmann weight for the two proteins to be separated by a distance $r$ is 
\be{
z(s_1,s_2,r)=e^{-H_{\mbox{\tiny non}}}\frac{2\pi}{A}(1+f_{12})r.
\label{zrd}
}\ee

\n
The probability that the proteins are separated by a distance $r$, regardless of their conformation, is
\be{
P(r)=\frac{Z(r)}{\left<Z\right>}=\frac{\sum_{s_1,s_2}z(s_1,s_2,r)}{\left<Z\right>},
}\ee

\n
from which we calculate the average separation
\begin{eqnarray}
\left<r\right> &=& \frac{1}{\left<Z\right>}\int Z(r)r\dd r \label{ravgformula}\\
&=&\frac{1}{\left<Z\right>}\sum_{s_1,s_2}e^{-H_{\mbox{\tiny non}}}\left(\delta\frac{\pi}{6}\sqrt{A}+\frac{2\pi}{A}\int_0^{\infty}f_{12}r^2\dd r\right) \nonumber.
\end{eqnarray}

\begin{figure}
\begin{center}
\includegraphics[width=3.5in]{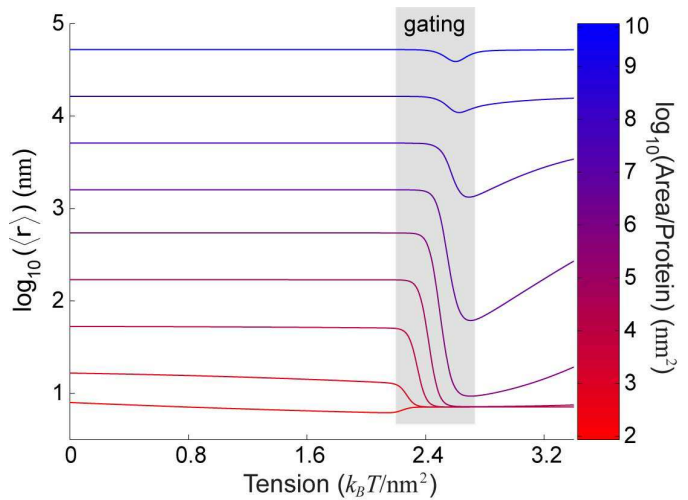}
\caption{  \small Average Separation between Proteins Drops Significantly due to Elastic Interactions. The average separation between two diffusing MscL proteins in a box of area $A$ is plotted as a function of tension for a range of areal densities, each shown as a different line color. The grey region roughly indicates when gating is occurring.  At low areal density (most blue) the conformational change does not draw the proteins significantly closer together.  As the areal density increases, the conformational change is able to draw the proteins up to $\sim100$ times closer than they would otherwise be.  At the highest areal density (most red) the steric constraint of available area intrinsically positions the proteins close to one another regardless of their conformation.  The average separation begins to increase again as higher tension weakens the open-open interaction.}
\label{Fig5}
\end{center}
\end{figure}

\begin{figure}
\begin{center}
\includegraphics[width=3.3in]{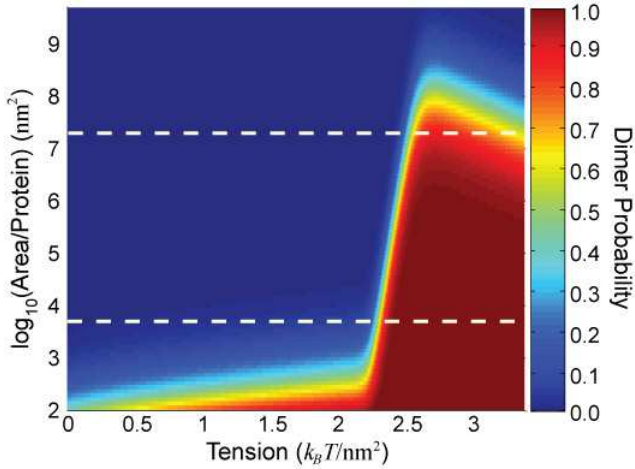}
\caption{\small Elastic Interactions Tightly Couple Conformational Change with Protein Dimerization. Diffusing MscL proteins are considered dimerized when they are close enough that they attract with an energy greater than $k_BT$.  At low area per protein, the net attractive closed-closed interaction is sufficient to dimerize the two channels part of the time.  As the area per protein increases, the closed-closed interaction is not strong enough to dimerize the two channels $-$ now dimerization only happens at higher tensions after both channels have switched to the open conformation. As the area per protein grows even larger, the open-open interaction is no longer strong enough to overcome entropy.  This loss of dimerization is amplified by the fact that the open-open interaction is weaker at higher tensions (see Figure \ref{Fig2}). The white dashed lines roughly indicate the range of areal densities for which dimerization probability and open channel probability are equal to each other (see Figure \ref{Fig4}).}
\label{Fig6}
\end{center}
\end{figure}

\n
This equation is valid as long as the area does not confine the proteins so severely that they are sterically forced to interact.  The constant $\delta$ is an order-one quantity that depends on the actual shape of the surface\footnote{On a surface, $\mathcal{S}(A)$, the average separation has an entropic component given by  $\int\int_{\mathcal{S}(A)}|{\bf r}_1-{\bf r}_2|\frac{\dd^2{\bf r}_1}{A} \frac{\dd^2{\bf r}_2}{A}=\delta\frac{\pi}{6}\sqrt{A}$.  The shape of the surface applies non-trivial bounds on this integral. }; for a square box, $\delta\simeq1$, and for a circle, $\delta\simeq\sqrt{2}$.  The average separation of two MscL proteins as a function of tension is plotted for various areal densities in Figure \ref{Fig5}.  Note that for certain densities, elastic interactions couple the conformational change from the closed to open state with a decrease in the average separation by more than two orders of magnitude.  Our estimates of biological membranes yield fairly high membrane-protein densities ($\sim100-1000\,\mbox{nm}^2$/protein) \cite{Engelman2004} which corresponds to the most highly confined conditions on the Figures \ref{Fig4}, \ref{Fig5} and \ref{Fig6}.  In the native {\it E. coli} plasma membrane, MscL, with a copy number of $\sim5$ \cite{Stokes2003}, is present at a density of $\sim10^5-10^6\,\mbox{nm}^2$/protein, which means that even membrane proteins expressed at a low level are subject to the effects of elastic interactions.  

To quantify the effects of interaction on the spatial organization of two channels, we define a `dimerized' state by the maximum separation below which two channels will favorably interact with an energy greater than $k_BT$ ({\it i.e.} $H_{\mbox{\tiny int}}(s_1,s_2,\tau,r)<-1$).  This defines a critical separation, $r_{c}(s_1,s_2,\tau)$, which depends on the conformations of each protein and the tension.  The probability that the two proteins are found with a separation less than or equal to $r_c$ is
\be{
P_{\mbox{\tiny dimer}}(\tau,\alpha)=\frac{1}{\left<Z\right>}\sum_{s_1,s_2}e^{-H_{\mbox{\tiny non}}}\left(\frac{\pi r_c^2}{A}+\frac{2\pi}{A}\int_0^{r_c}f_{12}r\dd r\right).
}\ee

\n
This `dimerization probability' is plotted as a function of tension and areal density in Figure \ref{Fig6}. 

At low tension and low area per protein, the channels are closed and near enough that the closed-closed interaction can dimerize them a fraction of the time.  Keeping the area per protein low, increasing tension strengthens the closed-closed interaction and the dimerization probability increases until tension switches the channels to the open state, where the significantly stronger open-open interaction dimerizes them essentially 100 percent of the time.  When the area per protein increases to moderate levels, as denoted by the white dashed lines in Figure \ref{Fig6}, the dimerization is strongly correlated with the conformational change to the open state.  The zero tension separation between the two proteins for this one-to-one correlation is $\sim40\,\mbox{nm}$ to $\sim2.2\,\mu\mbox{m}$.  Finally, when the area per protein is very large, entropy dominates, and neither the closed-closed, nor the open-open interaction is strong enough to dimerize the channels.

In summary, we have shown that over a broad range, areal density plays a non-trivial role in allowing two channels to communicate conformational information. This communication can lead to large changes in the average separation between two proteins and the probability that they will be found together in a dimerized state.  This may have implications for how conformational changes of transmembrane proteins in biological membranes are able to facilitate the formation of functional groups of specific proteins. \\

\n
{\bf\large Discussion}

In this section, we will perform a brief survey of other bilayer-mediated forces between proteins and make a comparison of their relative length and energy scales.  We will also address some of the finer details of our model and how boundary conditions can affect deformation energy around a protein.  Finally, we will suggest experiments using MscL to observe the predicted changes in conformational statistics, as well as provide evidence from previous experiments that leaflet interactions lead to significant changes in conformational statistics.  

  There are at least two other classes of purely bilayer-mediated forces between membrane proteins.  The first is a different type of bilayer deformation that bends the mid-plane of the bilayer.  This arises from transmembrane proteins with a trapezoidal shape that impose a bilayer slope at the protein-lipid interface \cite{Dan1998,OsterBJ2001}.  If the protein does not deform the bilayer too severely, the mid-plane deformation energy of a bilayer is 
\be{
G_{\mbox{\tiny mid}}=\int\left(\frac{\tau}{2}(\nabla h({\bf r}))^2+\frac{\kappa_b}{2}(\nabla^2 h({\bf r}))^2\right)\dd^2{\bf r},
}\ee

\n
where $h({\bf r})$ is the deviation of the height of the mid-plane from a flat configuration \cite{Wiggins2005,Helfrich1973}.  These kinds of interactions have been calculated for a variety of bilayer curvature environments and protein shapes at zero tension \cite{OsterBJ2001}.  Using a bilayer bending modulus of $\sim100\,k_BT$, attractive interactions of order $\sim1-5\,k_BT$ were found when the proteins were separated by 1-2 protein radii (which we estimate to be $5-10\,\mbox{nm}$ measured center-to-center for a typical transmembrane protein).  If we adjust the energy scale to be consistent with a PC bilayer bending modulus of $\sim14\,k_BT$ this lowers the interaction energetics to $\sim0.4-2\,k_BT$. Hence, although the length scale of appreciable interaction for mid-plane deformation is longer than for leaflet deformation, the interaction energies from leaflet deformation can be 10 times greater depending on protein geometry. The deformation fields $h({\bf r})$ and $u({\bf r})$ exert their effects independent of one another \cite{Wiggins2005}, suggesting that while energetically weaker than leaflet deformation, mid-plane deformation probably also contributes to the spatial organization and conformational communication between transmembrane proteins.  However, for the resting tension of many biological membranes \cite{Morris2001}, the interaction due to midplane deformation has a length-scale ($\sqrt{\kappa_b/\tau}\simeq50\,\mbox{nm}$) longer than the nominal spacing of proteins ($\simeq10-30\,\mbox{nm}$ \cite{Engelman2004}).  Thus, one protein can shield other proteins from feeling the deformation of a neighboring protein, and hence interactions are not (in general) pairwise additive.  In fact, this is a general feature for both leaflet and midplane elastic interactions -  they can be shielded by the presence of other proteins, and non-specific protein interactions can couple to conformation and position within the membrane in the same manner as the specific interactions we have explored in the previous sections.

The second class of bilayer-mediated forces is a product of the thermal fluctuations of the bilayer.  There is a small thermal force due to the excluded volume between two proteins, calculated via Monte Carlo methods to have a favorable $\sim2\,k_BT$ interaction \cite{Sintes1997}.  This force only exists when the proteins are separated by a fraction of the width of a lipid molecule.  There is also a long-range thermal force, due to the surface fluctuations of the bilayer, which tends to drive two rigid proteins closer together \cite{Goulian1993,ParkJFr1996}.  This force is proportional to $1/r^4$ and is generally attractive.  Estimates using this power law indicate that the interaction is $\sim1\,k_BT$ when the center-to-center separation is roughly 2 protein radii.  Though elegant, the derivation of this force is only valid in the far-field, thus how this force might contribute to conformational communication between proteins in close proximity is not entirely clear. 

To gauge the overall importance of leaflet interactions, the virial coefficient used in Eq. \ref{eqV},
\be{
C_{\tiny V}=2\pi\int_0^{\infty}f_{12}(r)r\dd r,
}\ee

\n
quantifies how the combination of length and energy scales leads to a deviation from non-interacting behavior; it is {\it exponentially} sensitive to the energy but only {\it quadratically} sensitive to the length-scale. One can interpret the virial coefficient as the area per particle that makes the competing effects of entropy and interaction equivalent. Using this measure, we estimated the virial coefficients for all of these bilayer-mediated forces and found that leaflet deformations, while having a short length scale, actually lead to the most significant deviation from non-interacting behavior, due to their high energy scale.  We estimate the virial coefficients from leaflet interactions to be $\sim10^4-10^6\,\mbox{nm}^2$, while mid-plane bending interactions are $\sim10^3\,\mbox{nm}^2$, and the thermal forces $\sim10^2\,\mbox{nm}^2$.

Examining our elastic model in greater detail, we have assumed that the slope of the leaflet at the protein-lipid interface is zero, which eliminates any dependence on the spontaneous and Gaussian curvatures of the leaflet.   In a more general continuum-mechanical theory, the slope would be left as a free parameter with respect to which the energy could be minimized \cite{AndersenBJ1998}.  We examined this possibility and found that, at most, the energy was reduced by a factor of two. Spontaneous curvature couples to the slope of the leaflet at the protein-lipid interface, however the spontaneous curvature of bilayer forming lipids, such as phosphatidylcholines, is small \cite{Boal}.  In addition, for proteins whose radius is larger than $\lambda$, if we assume the modulus associated with Gaussian curvature is of the same magnitude as the mean curvature modulus ($\kappa_b$) \cite{KozlovBJ2004}, the Gaussian contribution to the deformation energy is a second-order effect.  We also examined the possibility of a term proportional to $(\nabla u)^2$; using the interfacial tension ($\sim5\,k_BT/\mbox{nm}^2$) as a modulus for this term; these effects were also second-order.  Finally, we imposed the `strong hydrophobic matching' condition at the protein-lipid interface, assuming that the interaction of lipids with the hydrophobic zone of the protein is very favorable.  Relaxing this condition would result in a decrease in the magnitude of the hydrophobic matching condition, $u_o$, and hence an overall decrease in energetics \cite{Wiggins2005}.

There are also experimental and mechanical reasons to believe the boundary slope on a cylindrical protein is small.  The membrane protein gramicidin was used to comment on this so-called `contact angle' problem of lipid-protein boundary conditions \cite{Huang1986,Elliott1983}. It was found that indeed the slope was nearly zero.  From a mechanical standpoint, if the lipids are incompressible, a positive boundary slope that deviates significantly from zero would correspond to the creation of an energetically costly void at the protein-lipid interface when the protein is shorter than the bilayer.  Conversely, lipid would have to penetrate the core of the protein to produce a negative slope when the protein is taller than the bilayer, again a very costly proposition.

We examined a roughly cylindrical protein and demonstrated the interesting effects elastic interactions would have in such cases. However, the scope of possible effects increases when non-cylindrical proteins are considered.  Most notably, non-cylindrical cross-sections allow for orientational degrees of freedom in the interaction, hence such proteins do not just attract or repel each other, but would have preferred orientations in the membrane with respect to each other.

Measuring the changes in conformational statistics of two MscL proteins held at a fixed separation would allow for quantitative verification of our predictions. Electrophysiology is a common tool used to probe the conformation of ion channels, and is routinely used to measure the open probability of a single MscL protein {\it in vitro} \cite{SukharevJGP1999,SukharevBJ2004,PerozoNSB2002}.  Cysteine point mutations on the outer edges of two MscL proteins \cite{PerozoJGP2001} could be covalently linked \cite{Karlin1998,Karlin2001,Javitch2003,Dowhan2005} by a polymer with a specific length ($\sim0.5-10\,\mbox{nm}$) to control the separation distance \cite{Haselgrubler1995,Blaustein2000}.   Linking stoichiometry could be controlled genetically \cite{SukharevJMB1999} to ensure one channel interacts with only one other channel.

Similar experiments have been performed using gramicidin $A$ channels \cite{Goforth2003}.  The conducting form of gramicidin $A$ is a cylindrical transmembrane protein which, like MscL, tends to compress the surrounding bilayer \cite{Huang1986,HuangBJ1999,AndersenBJ1999} and hence have a beneficial interaction. Electrophysiology of polypeptide-linked gramicidin channels \cite{Goforth2003} qualitatively supports our hypothesis that the beneficial interaction of the deformed lipids around two gramicidin channels significantly increases the lifetime of the conducting state. As another example, recent FRET studies showed that oligomerization of rhodopsin is driven by precisely these kinds of elastic interactions, and exhibits a marked dependence on the severity of the deformation as modulated by bilayer thickness \cite{Botelho2006}.  Additionally, recent experimental work has shown that the bacterial potassium channel KscA exhibits coupled gating and  spatial clustering in artificial membranes \cite{Molina2006}.

In summary, we have demonstrated that leaflet deformations are one of the key mechanisms of bilayer-mediated protein-protein interactions.  We provided support for our choice of boundary conditions at the protein-lipid interface, and suggested that extensions of our model have exciting possibilities for the specificity of elastic interaction. Finally, we suggested how one might measure the predicted changes in conformational statistics and drew an analogy to previous gramicidin channel experiments.\\

\n
{\bf\large Conclusion}

  We have described the important role of an elastic bilayer in the function of, and communication between, membrane proteins.  Over a wide range of areal densities, transmembrane proteins can communicate information about their conformational state via the deformations they cause in the surrounding bilayer.  We demonstrated with a model protein, the tension-sensitive channel MscL, how deformations lead to elastic forces and result in cooperative channel gating.  Additionally, we found that elastic interactions strongly correlate conformational changes to changes in spatial organization, aggregating two channels even at low areal densities, bringing them together over very large distances relative to their size.

The elastic theory presented here can be easily expanded to include more complex deformation effects (such as spontaneous curvature) and protein shapes.  Our calculations for the conformational statistics, average separation, and dimerization are insensitive to the actual stimulus triggering the conformational change.  Hence, we suggest that elastic interactions are likely to play a role in the function and organization of many membrane proteins which respond to environmental stimuli by forming functional groups of multiple membrane proteins. Recent work suggests chemotactic receptors in {\it E. coli} function by precisely this kind of spatially clustered and conformationally coupled modality \cite{Skoge2006}.\\

\n
{\bf\large Materials and Methods}

{\small 
To compute the pairwise elastic potentials in Figure \ref{Fig2}, we discretize  
the bilayer height, $\eta(\rho)$, and minimize the deformation  
energy in Eq.~\ref{eqG} using a preconditioned conjugate gradient approach.  
A separate minimization with the aforementioned boundary conditions, including the zero-slope boundary condition, was computed  
for each combination of channel configurations, protein-protein  
separation, and bilayer tension. Except in the regions of the  
bilayer nearby a protein at position $(x_o,y_o)$, we use a Cartesian  
grid with spacing $dx=dy=0.1\lambda=0.093\,\mbox{nm}$.  
However, since deformations in the bilayer are largest at the  
circular membrane-protein interface, we interpolate between a polar  
grid at the interface at ${\bf r}=r_o$ and a Cartesian grid along the  
square $\cal S$ defined by $|x-x_o|<\Delta,|y-y_o|<\Delta$, where $\Delta$ is chosen to be an integral multiple of $dx$. This  
interpolation ensures an accurate estimate of the elastic deformation  
energy of a single protein and preserves the symmetry of the protein  
in its immediate vicinity.

The lines connecting the grid points along $\cal S$ define $n_\theta$  
angular grid points $\theta_i$ $(i=1,\ldots,n_\theta)$, and $n_r+1$  
grid points within the interpolation region are defined by the polar  
coordinates $(r_{ij},\theta_i)=(r_o+\delta r_ij/n_r,\theta_i)$, where  
$r_o$ is the radius of the protein and the distance from the center  
of the protein to $\cal S$ along $\theta_i$ is $r_o+\delta r_{i}$ (\emph
{e.g.}, for $\theta_i=0$, $\delta r_i=\Delta-r_o$; for $\theta_i=\pi/4
$, $\delta r_i=\Delta\sqrt{2}-r_o$. For a protein in the open or  
closed configuration, $\Delta$ was chosen such that $n_\theta=320$ or  
$224$, respectively.

The deformation energy determined using this numerical relaxation  
method is converged with respect to $dx$, $\Delta$ and the overall  
dimensions of the bilayer ($18.5\,\mbox{nm}\times 37.1\,\mbox{nm}$), and reproduces  
the analytic results for a single protein given by Eq. \ref{eqsingle}. The  
elastic potentials were determined over the relevant range of channel  
separations from 0 to $\sim 8\,\mbox{nm}$ (measured from protein edge to  
protein edge), and for a range of bilayer tensions from 0 to $3.4\,k_BT/\mbox{nm}^2$.\\
}

\n
{\bf\large Supporting Information}\\
\n
{\small
{\bf Accession Numbers}\\
\n
The primary accession numbers (in parentheses) from the Protein Data Bank (http://www.pdb.org) are:  mechanosensitive channel of large conductance (2OAR; formerly 1MSL), gramicidin A ion channel (1GRM), bacterial potassium ion channel KscA (1F6G), and bovine rhodopsin (1GZM).
  }\\

\n
{\bf\large Acknowledgments}

{\small
We would like to thank Doug Rees, Olaf Andersen, Pierre Sens, Sergei Sukharev, Nily Dan, Jennifer Stockdill, and Ned Wingreen for their thoughtful comments on the manuscript, Chris Gandhi for his input into possible experiments and Ben Freund for useful discussion.  

{\bf Author contributions.} TU conceived of the experiment and performed the analytical calculations.  TU, EP and KH performed numerical simulations.  TU, KH, EP and RP analyzed the data and wrote the paper.

{\bf Funding.}  RP acknowledges the support of the National Science Foundation Award No.~CMS-0301657.  TU and RP acknowledge the support of the NSF CIMMS Award No.~ACI-0204932 and NIRT Award No.~CMS-0404031 as well as the National Institutes of Health Director's Pioneer Award.~EP was supported by the Department of Homeland Security Graduate Fellowship program and the NIH Director's Pioneer Award.  KH was supported by the NIH Award No.~A1K25 GM75000.  Part of this work took place at the Kavli Institute for Theoretical Physics, Santa Barbara, CA and the Aspen Center for Physics, Aspen, CO.
}

{\footnotesize

}

\end{document}